\documentclass[prl,reprint,noshowpacs,superscriptaddress,floatfix,letterpaper,notitlepage]{revtex4-1}
\usepackage{amsmath,amssymb,amsbsy,amsfonts,amsthm,bbm,bm,mathtools,mathrsfs}
\usepackage{color}
\usepackage{physics}
\usepackage[dvipsnames]{xcolor}
\usepackage[colorlinks=true,citecolor=Cerulean,linkcolor=RubineRed,urlcolor=Cerulean]{hyperref}
\usepackage{empheq}
\usepackage{stackengine}
\usepackage[normalem]{ulem}

\usepackage{lettrine}
\usepackage{setspace}

\definecolor{rabPurple}{RGB}{132,104,201}

\newcommand{\speedtime}{\tau_{\!\scriptscriptstyle\mathcal{A}}}

\def\date{}

\begin{document}

\title{Time-information uncertainty relations in thermodynamics}

\author{Schuyler~B.~Nicholson}
\affiliation{Department of Chemistry,\
 University of Massachusetts Boston,\
 Boston, MA 02125
}
\affiliation{Center for Quantum and Nonequilibrium Systems,\
  University of Massachusetts Boston,\
  Boston, MA 02125
}

\author{Luis~Pedro Garc\'{i}a-Pintos}
\affiliation{Department of Physics,\
  University of Massachusetts Boston,\
  Boston, MA 02125
}
\affiliation{Joint Center for Quantum Information and Computer Science,\
  NIST/University of Maryland,\
  College Park, Maryland 20742}
\affiliation{Joint Quantum Institute,\
  NIST/University of Maryland,\
  College Park, Maryland 20742}

\author{Adolfo del Campo}
\affiliation{Donostia International Physics Center, E-20018 San Sebasti\'an, Spain}
\affiliation{IKERBASQUE, Basque Foundation for Science, E-48013 Bilbao, Spain}
\affiliation{Department of Physics,\
  University of Massachusetts Boston,\
  Boston, MA 02125
}

\author{Jason~R.~Green}
\email[]{jason.green@umb.edu}
\affiliation{Department of Chemistry,\
  University of Massachusetts Boston,\
  Boston, MA 02125
}
\affiliation{Center for Quantum and Nonequilibrium Systems,\
  University of Massachusetts Boston,\
  Boston, MA 02125
}
\affiliation{Department of Physics,\
  University of Massachusetts Boston,\
  Boston, MA 02125
}

\begin{abstract}

Physical systems that power motion and create structure in a fixed amount of
time dissipate energy and produce entropy. Whether living or synthetic, systems
performing these dynamic functions must balance dissipation and speed. Here, we
show that rates of energy and entropy exchange are subject to a speed limit --
a time-information uncertainty relation -- imposed by the rates of change in
the information content of the system. This uncertainty relation bounds the
time that elapses before the change in a thermodynamic quantity has the same
magnitude as its initial standard deviation. From this general bound, we
establish a family of speed limits for heat, work, entropy production, and
entropy flow depending on the experimental constraints on the system. In all of
these inequalities, the time scale of transient dynamical fluctuations is
universally bounded by the Fisher information. Moreover, they all have a
mathematical form that mirrors the Mandelstam-Tamm version of the time-energy
uncertainty relation in quantum mechanics. These bounds on the speed of
arbitrary observables apply to transient systems away from thermodynamic
equilibrium, independent of the physical assumptions about the stochastic
dynamics or their function.

\end{abstract}

\maketitle

\lettrine[lraise=0.1, nindent=0em, slope=-.5em]{M}{any} problems in science and
engineering involve understanding how quickly a physical system transitions
between distinguishable states and the energetic costs of advancing at a given
speed. While theories such as thermodynamics and quantum mechanics put
fundamental bounds on the dynamical evolution of physical systems, the form and
function of the bounds differ. Take Clausius' version of the second law of
thermodynamics~\cite{Callen85}. It is an upper bound on the heat exchange in
traversing equilibrium thermodynamics states--an inequality that limits the
efficiency of heat engines without an explicit notion of time or fluctuations.
By contrast, Mandelstam and Tamm's version of the time-energy uncertainty
relation in quantum
mechanics~\cite{mandelstam1945uncertainty,margolus1998maximum} is a limit on
the speed at which quantum systems can evolve between two distinguishable
quantum states. Given this important, and longstanding, contrast between these
two pillars of physics, we explore thermodynamic bounds that are analogous to
those in quantum mechanics, bounds that are independent of the system
dynamics~\cite{pietzonka2018universal,gingrich2016dissipation,nicholson2018nonequilibrium,barato2018bounds,dechant2018multidimensional}
and set limits on the speed of energy and entropy exchange.

Thermodynamic uncertainty
relations~\cite{uffink1999thermodynamic,schlogl1988thermodynamic} have been
found where fluctuations in dynamical currents are bounded by the entropy
production
rate~\cite{barato2015thermodynamic,gingrich2016dissipation,pietzonka2017finite,maes2017frenetic}.
These relations apply to small systems and are part of stochastic
thermodynamics~\cite{Jarzynski11, seifert2012stochastic, Vandenbroeck2015,
boyd2016identifying, sagawa2013information}, a framework in which thermodynamic
quantities, such as heat, work, and entropy, can be treated at the level of
individual, fluctuating trajectories. In parallel to these discoveries, there
have been advances in quantum-mechanical uncertainty relations or ``speed
limits'' that constrain the speed at which dynamical variables evolve.  They
employ the mean~\cite{margolus1998maximum},
variance~\cite{mandelstam1945uncertainty}, or higher-order moments of the
energy~\cite{Zych06,Margolus11}. These quantum speed limits have recently been
generalized to open quantum systems embedded in an
environment~\cite{Taddei13,delcampo13,DeffnerLutz13,Luispe19}, paving the way
to their application in the classical domain. The existence of speed limits,
regardless of the classical or quantum nature of the system, was first pointed
out by Margolus~\cite{Margolus11}. Only recently have analogous bounds been
established in classical systems and applied to Liouville dynamics in phase
space~\cite{Shanahan18,Okuyama18} (also see related
work~\cite{Takahashi16,Shiraishi18}). While there has been rapid progress on
thermodynamic uncertainty relations~\cite{horowitz2019thermodynamic}, it
remains to be seen whether there are speed limits in thermodynamics whose
generality rivals those in quantum mechanics.

What governs the speed at which heat, work, and entropy are exchanged between a
system and its surroundings? Is there a universal quantity that bounds the
speed at which thermodynamic observables evolve away from equilibrium?
Motivated by these questions, we derive a family of limits to the speed with
which a system can pass between nonequilibrium states and the heat, work, and
entropy exchanged in the process.

\smallskip

\noindent{\textbf{\textsf{Equation of motion for thermodynamic observables}}}\\
\noindent Let us consider a generic classical, physical system operating
irreversibly, out of thermodynamic equilibrium. The stimulus for the time
evolution of the physical system can be the removal of a constraint or the
manipulation of an experimental control parameter $\lambda$, such as
temperature or volume. As is common in stochastic
thermodynamics~\cite{seifert2012stochastic}, we adopt a mesoscopic description
and take the system to have a finite number of configurations $x = 1,2,\ldots,
N$ with initial probability $p_x(t_0)$. As currents in energy and matter cause
the system to evolve, the probability distribution will generally differ from
that of a Gibbsian ensemble. Our working assumption is that the dynamical
evolution smoothly transforms the probability, $p_x[\lambda(t)] = p_x(t) =
p_x$, of each state $x$ at time $t$ with a rate $\dot{p}_x =
dp_x/dt$~\footnote{All quantities are time dependent unless explicitly
indicated otherwise.}.

During the nonequilibrium process, experimental measurements of an observable
$A$ for this classical system correspond to  time-dependent statistical moments
$\left<A^n\right> = \sum_x^N p_x a_x^n$ of the configuration observables
$a_x(t)=a_x$. The Shannon entropy~\cite{Shannon}, for example, is the
expectation value of the surprisal $I_x := -\ln p_x$, which measures the
information gained by observing the system in state $x$. With these minimal
specifications, our first main result is that the ensemble average of any
observable $A$ obeys an equation of motion,
\begin{equation}
  \frac{d\langle A\rangle}{dt} = -\operatorname{cov}(A,\dot{I}) + \left<\frac{dA}{dt}\right>.
  \label{eq:evolution}
\end{equation}
The covariance measures the amount of linear correlation between $A$ and the
surprisal rate $\dot{I}_x=dI_x/dt$,
\begin{equation}
  -\dot{\mathcal{A}} := \operatorname{cov}(\dot{I},A)= \big\langle (A - \langle A \rangle ) (\dot{I} - \langle\dot{I}\rangle) \big\rangle.
\end{equation}
This evolution law makes no additional physical or modeling assumptions and
holds for general processes away from thermodynamic equilibrium (see proof in
Supplementary Material, SM. \textsf{I}).

Another form of the evolution equation for energy is well-known in stochastic
thermodynamics. For a system with a finite number of energy states,
$a_x=\epsilon_x$, it is the stochastic first law, $\dot{U} = d\langle
\epsilon\rangle/dt = \dot{Q} + \dot{W}$~\cite{Vandenbroeck2015}.  Comparing the
first law to our result gives a statistical representation for the flux of
work, $\langle d\epsilon/dt\rangle$, and the flux of heat, $\dot{Q} = \sum_x^N
\dot p_x\delta\epsilon_x = -\operatorname{cov}(\dot{I},\epsilon)$, where we
have shifted the energy scale: $\delta \epsilon_x := \epsilon_x - U$. Thus, we
find a new definition of energy exchanged between a system and its surroundings
as heat: heat flux is a measure of the linear correlation between energy and
information rates. While the covariance measures the linear relationship
between random variables, it applies even when $\dot{I}_x$ and $\epsilon_x$ are
nonlinearly related. Any observable satisfying  $\sum_x^N\dot{p}_xa_x$, such as
the entropy rate $\dot{S}$, can be expressed as this covariance.

The mathematical form of Eq.~(\ref{eq:evolution}) is strikingly similar to the
Ehrenfest theorem in quantum (classical) mechanics. Particularly important is
that the covariance fulfills the role of the mean commutator (Poisson bracket)
of a quantum (classical) mechanical observable and the quantum (classical)
Hamiltonian~\cite{Messiah61}. For this broad class of classical stochastic
systems, the surprisal rate, not the Hamiltonian, is the observable to which
all others are compared. Given this analogy, we explore whether other
relationships in quantum mechanics extend to classical, stochastic observables
built on fluctuations and uncertainty.

\smallskip

\begin{figure}
  \centering
  \includegraphics[width=0.8\columnwidth]{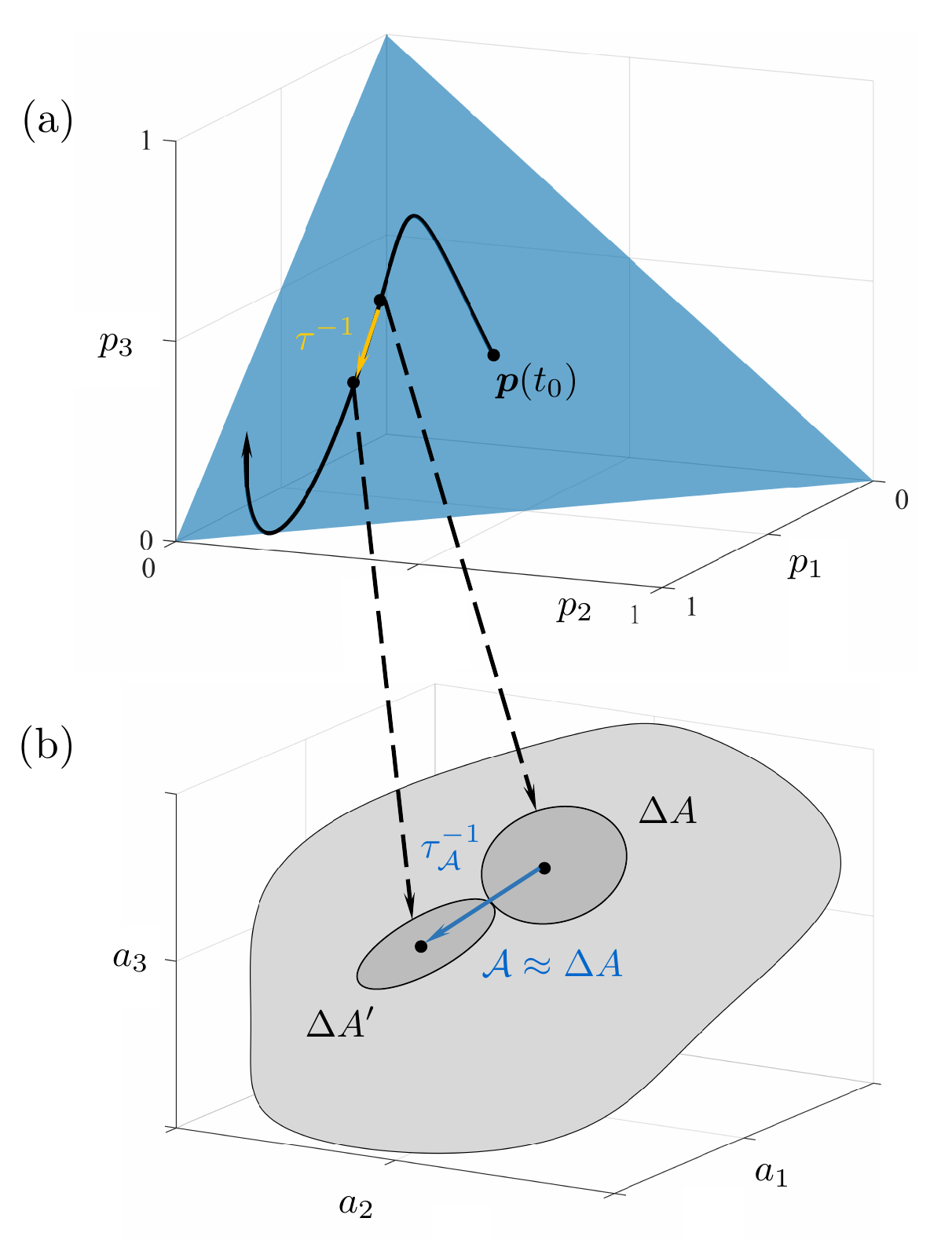}

  \caption{\label{fig:fig1} \textbf{Statistical distinguishability of
thermodynamic observables.} As a result of an (a) underlying stochastic
dynamics and probability distribution over configurations, an (b) observable
$A$ of a system evolves in time away from thermodynamic equilibrium.  The
distributions of $A$ at two times are distinguishable when their standard
deviations do not overlap. The time $\tau$ to reach a distinguishable state is
when the path function $\mathcal{A}$ has the magnitude of one initial standard
deviation $\Delta A$.}

\end{figure}

\noindent{\textbf{\textsf{Observable fluctuations and intrinsic speed}}}\\
\noindent As the nonequilibrium dynamics of the system unfold, observables
will fluctuate in time. Experimental measurements of these observables will
then deviate from their mean $\delta a_x = a_x-\left<A\right>$. The variance,
\begin{equation}
  (\Delta A)^2 = \sum_x^N p_x \delta a_x^2,
\end{equation}
is a measure of the uncertainty in the measurements. It also measures the
distinguishability between two states of the observable $A$,
Fig.~(\ref{fig:fig1}). Two observable states are distinguishable in the sense
used by Wootters~\cite{StatDistQuantum} if the statistical distance between
them is greater than their combined uncertainty: dist$(A,A') \geq \Delta A +
\Delta A'$ and $A \neq A'$.

From the equation of motion, changes in the state function $\Delta \langle
A\rangle$ are the result of two path functions. One dynamical measure of the
variation in $A$ is the amount of time for the magnitude of the path function
$\mathcal{A}=\int\dot{\mathcal{A}}\,dt$ to have the value of one standard
deviation $\Delta A$; heat, for example, is $Q=\int\dot{Q}\,dt$. This time is
approximately:
\begin{equation}
  |\mathcal{A}| = \left|\int_{t_{0}}^{t_{0}+\tau_{\!\scriptscriptstyle \mathcal{A}}}\dot{\mathcal{A}}\,dt\right|
  \approx |\dot{\mathcal{A}}|\,\tau_{\!\scriptscriptstyle \mathcal{A}} \approx \Delta A,
\end{equation}
which suggests the definition of the intrinsic speed,
\begin{equation}
  \frac{1}{\speedtime} := \frac{|\dot{\mathcal{A}}|}{\Delta A} = \frac{|\!\operatorname{cov}(\dot{I},A)|}{\Delta A},
\end{equation}
for any dynamical variable $A$. Speed limits for quantum-mechanical observables
define a time scale with a similar form~\cite{Messiah61}. However, in quantum
mechanics, the second term $\left<d\mathcal{A}_\textrm{QM}/dt\right>=0$ is zero
in the Ehrenfest equation of motion, which leads to the definition of a time
scale in terms of the expectation value of the commutator,
$\langle[\mathcal{A}_\textrm{QM},\mathcal{H}_\textrm{QM}]\rangle$, a role
played here by the covariance. Since, $\left\langle dA/dt\right\rangle$ does
not have to be zero here, we instead define the time scale in terms of the
covariance $\dot{\mathcal{A}} = -\operatorname{cov}(\dot{I},A)$.

\smallskip

\noindent{\textbf{\textsf{Information fluctuations and intrinsic speed}}}\\
\noindent The intrinsic speed of each observable $A$ measures its sensitivity
to changes in the distribution over configurations. To bound this speed for
generic observables, it is natural to examine the time scale for the
probability distribution to evolve to a distinguishable
state~\cite{SalamonB83,*SalamonNB85}. There is evidence in both
quantum~\cite{StatDistQuantum,DistQuantumTwo} and
classical~\cite{MLthermo,flynn2014measuring,*nichols2015order,heseltine2016novel}
settings that the square root of the Fisher information~\cite{FisherInfo},
$\sqrt{I_F}$, defines such a speed for time-varying probability distributions.
The Fisher information parameterized by time is a measure of distance $ds$
between neighboring distributions,
\begin{align}
  ds^2 &= \sum_{i,j} \frac{d\lambda^i}{dt}g_{ij}\frac{d\lambda^j}{dt}dt^2 = I_F \,dt^2,
  \label{eq:distance}
\end{align}
where the Fisher metric is
\begin{align}
  g_{ij} = \left\langle\frac{\partial \ln p_x}{\partial\lambda_i}\frac{\partial \ln p_x}{\partial\lambda_j}\right\rangle.
  \label{eq:IFMet}
\end{align}
This statistical distance can be interpreted as a measure of the
distinguishability between $p_x(t)$ and
$p_x(t+dt)$~\cite{heseltine2016novel,StatDistQuantum}. Looking at the physical
dimensions, $\sqrt{I_F}$ is a ``speed'' relating the dimensionless measure of
distance $ds$ and infinitesimal increment of time $dt$.

The Fisher information~\cite{FisherInfo} is also a measure of fluctuations. It
measures fluctuations in the surprisal rate, $\dot{I}_x = -d\ln p_x/dt$:
\begin{equation}
  I_F := \sum_{x}^N p_x\left(\frac{d\ln p_x}{dt}\right)^2 = \Delta {\dot{I}}^2.
\end{equation}
The surprisal rate fluctuates only for temporally-varying distributions, that
is, only in systems out of thermodynamic equilibrium. By the dimensional
analysis above, fluctuations in the information content, $1/\Delta\dot{I} =
1/\sqrt{I_F}$, set a time scale for the evolution of the probability
distribution in systems out of equilibrium, $\tau :=
1/\sqrt{I_F}$~\cite{flynn2014measuring,nichols2015order,kim2016geometric}.
But, does this time scale provide a general bound on the speed at which
nonequilibrium observables evolve?

\smallskip

\noindent{\textbf{\textsf{Time-information uncertainty relation}}}\\
\noindent In the quantum setting, the time variation of the mean value of an
observable is subject to the Mandelstam-Tamm time-energy uncertainty
relation~\cite{mandelstam1945uncertainty}. Just as in the quantum-mechanical
formalism, we can place bounds on the uncertainty in thermodynamic observables,
regardless of the dynamical variable or the stochastic dynamical law governing
the probability distribution over configurations. The fluctuations in $A$ and
$\dot{I}$ upper bound their covariance through the inequality:
\begin{equation}
  \label{eq:generalbound}
  |\dot{\mathcal{A}}| = |\!\operatorname{cov}(\dot{I},A)| \leq \Delta\dot{I} \Delta A.
\end{equation}
This bound is an uncertainty relation for any observable $A$, set by
fluctuations in the surprisal rate, $\Delta\dot{I} = \sqrt{I_F}$. Examples we
will explore here include the fluxes of heat $\dot{Q}$, dissipated work
$\dot{W}_\textrm{diss}$, system entropy $\dot{S}$, entropy production
$\dot{S}_i$, and entropy flow $\dot{S}_e$.

As an immediate consequence of this uncertainty relation, fluctuations in the
surprisal rate are an upper bound on the speed of any dynamical variable,
\begin{equation}
  \Delta\dot{I} \geq 1/\speedtime.
  \label{eq:TimeInq}
\end{equation}
Clearly, the Fisher information is intimately connected to the stochastic
dynamics over configurations. And, in this classical uncertainty relation,
$\speedtime\Delta\dot{I} \geq 1$, it sets the time scale that bounds the time
scale of all other dynamical quantities: A system out of thermodynamic
equilibrium with a spread $\Delta\dot{I}=\sqrt{I_F}$ in surprisal rate takes a
time of at least $\speedtime \geq 1/\Delta\dot{I}$ for the path function
$\mathcal{A}$ to have the value of one standard deviation $\Delta A$. This
time-information uncertainty relation assumes a differentiable distribution
over a finite number of discrete states but it makes no model assumptions about
the stochastic dynamics, the proximity to equilibrium, the size of the system,
or the protocol driving the system out of equilibrium.

In this time-information uncertainty relation, as in the Mandelstam-Tamm's
version of the time-energy uncertainty relation, $\speedtime$ is the amount of
time that elapses before $\mathcal{A}$ changes by one standard deviation in
$A$. The inequality represents a bound that the spread in surprisal rates
places on the time scale of measurable changes in the dynamical variables.  The
more concentrated the surprisal rates, the slower any observable $A$ will
change in time. For example, when the system is at equilibrium or in a
nonequilibrium steady-state with a finite $\Delta A$, the uncertainty in the
surprisal rate vanishes, $\Delta \dot{I} = 0$. No matter what $A$ is being
considered, $|\dot{\mathcal{A}}| = 0$ and the time scale $\speedtime$ is
infinite. Conversely, if any observable exhibits a rapid variation with time,
then the distribution must have large fluctuations in the surprisal rates and a
large Fisher information. For example, when the system is driven quickly
through nonequilibrium states so that $|\dot{\mathcal{A}}|\to\infty$, the time
scale $\speedtime \to 0$.  Accomplishing such an extreme change in the mean
requires a corresponding change in the distribution and
$\Delta\dot{I}\to\infty$.

These general results have particular physical significance within
thermodynamics. Thermodynamics can be seen as having specific representations
depending upon the experimental conditions, which set the natural variables and
appropriate thermodynamic potential~\cite{Callen85}. Here, we focus on the most
fundamental representations and establish a family of time-information
uncertainty relations.

\smallskip

\noindent\textbf{\textsf{Thermodynamic observables}}\\
\noindent\textbf{\textsf{\small I.\@ Energy.}}\\
\noindent Energy transfer between a system and its surroundings can be divided
into heat and work. The internal energy of a macrostate is $U = \sum_x^N
p_x\,\epsilon_x$, where $\epsilon_x$ is the energy of state $x$. As we have
seen, the stochastic first law~\cite{Vandenbroeck2015}, $\dot{U} = \dot{Q} +
\dot{W}$, has the form of the evolution equation Eq.~(\ref{eq:evolution}). The
corresponding uncertainty relationship upper bounding the rate of heat exchange
is:
\begin{equation}
  |\dot{Q}| = |\!\operatorname{cov}(\dot{I},\epsilon)| \leq \Delta {\dot{I}} \, \Delta {\epsilon},
  \label{eq:flucBnd}
\end{equation}
where $\Delta \dot{I}$ and $\Delta \epsilon$ are the standard deviations in the
surprisal rate and energy, respectively. At equilibrium, where $\dot{Q}=0$ and
$\dot{p}_x = 0$ $\forall x$, the bound is trivially saturated. Away from
stationary states, the product of information rate and energy fluctuations
limit the rate at which energy can be absorbed or dissipated as heat.

If there is no work done on or by the system, then these fluctuations also
bound the internal energy flux $\dot{U} = \dot{Q}$ and the time scale is the
time required for the internal energy to change by a standard deviation in the
energy fluctuations: $\Delta U = \Delta\langle\epsilon\rangle =
\Delta\epsilon$.

If the energy fluctuations are fixed and of order $k_BT$, then the speed
$\tau_{\scriptscriptstyle Q}^{-1} = |\dot{Q}|/k_BT$ is bounded by the
fluctuations in surprisal rate, $\beta|\dot{Q}| \leq \Delta {\dot{I}}$. So, on
the one hand, the more concentrated the surprisal rates are around a given
value during the nonequilibrium process, the slower the maximum rate at which
energy is exchanged as heat. On the other hand, higher rates of heat exchange
are attainable only at the expense of a broader distribution of surprisal
rates. Or, to put it simply, fluctuations in the surprisal rate constrain
nonequilibrium heat flow.

\smallskip

\noindent\textbf{\textsf{\small II.\@ Entropy.}}\\
\noindent An uncertainty relation for entropy exchange complements that for
energy exchanged as heat. As the ensemble average of the surprisal, the Shannon
entropy $S/k_B = -\sum_x^N p_x\ln p_x$ also satisfies the equation of motion,
Eq.~(\ref{eq:evolution}).  Only the covariance term survives and the equation
of motion becomes (SM.~\textsf{VI}):
\begin{equation}
  \dot{S}/k_B = -\sum_x^N \dot{p}_x \ln p_x
  = -\operatorname{cov}(\dot{I},I).
  \label{eq:CovEntropy}
\end{equation}
The rate of change of the entropy measures the linear correlation between the
surprisal and its speed. Thus, the entropy rate, 
\begin{align}
  |\dot{S}|/k_B \leq \Delta {\dot{I}} \, \Delta{I},
  \label{eq:EntropyFluc}
\end{align}
is bounded by the spread in information-theoretic quantities, the surprisal and
its rate of change. The intrinsic time scale $\tau_{\scriptscriptstyle S} =
k_B\Delta I/|\dot{S}|\leq 1/\Delta\dot{I}$ measures the time needed for the
Shannon entropy to change by one standard deviation in the surprisal
fluctuations.

A common approach in nonequilibrium thermodynamics~\cite{deGrootM84} is to
divide the rate of entropy change for the system into the rate of entropy
production internal to the system $\dot{S}_i$ (the irreversible entropy
production rate) and the rate of entropy exchanged with the surroundings
$\dot{S}_e$ (the entropy flow rate): $\dot S = \dot S_i + \dot S_e$. Given the
Clausius inequality~\cite{Callen85} and the positivity of the entropy
production, $\dot S_i \geq 0$, it follows that $|\dot S_e|/k_B \leq \Delta
{\dot{I}} \, \Delta{I}$ and $\dot S_i/k_B \leq \Delta {\dot{I}} \, \Delta{I}$.

\smallskip

\noindent\textbf{\textsf{\small III.\@ Dissipated work.}}\\
\noindent The tendency of physical systems to increase entropy can be harnessed
to do useful work. However, unless the process is thermodynamically reversible,
some energy will be dissipated. For a system in contact with a heat bath at
fixed temperature $T=1/k_B\beta$, the nonequilibrium free energy is $F = U -
\beta^{-1} S$~\cite{seifert2012stochastic}. The rate of dissipated work or
dissipated power, $\dot{W}_\textrm{diss} = \dot W - \dot F$, also satisfies a
time-information uncertainty relation.

Using our results for the fluxes of heat and entropy, the dissipated power,
\begin{equation}
  \beta \dot{W}_\textrm{diss} = \beta\operatorname{cov}(\dot I,\epsilon) - \operatorname{cov}(\dot I,I),
  \label{eq:Wdiss}
\end{equation}
is the difference in the linear correlation of the information and the energy
with $\dot I$. The
time-information uncertainty relation is found using the triangle inequality
and the Clausius inequality $|\dot{S}|/k_B \geq \beta|\dot{Q}|$:
\begin{align}
  \beta |\dot{W}_\textrm{diss}| &\leq |\dot S|/k_B + \beta |\dot Q|
  \leq 2\Delta \dot I\Delta I.
\end{align}
Again, the rate of change in the information content of the distribution is the
reference for a thermodynamic observable.

\smallskip

\noindent{\textbf{\textsf{Saturation of the uncertainty relation}}}\\
\noindent A sufficient condition to saturate the uncertainty relation in
Eq.~\eqref{eq:generalbound}, $|\dot{\mathcal{A}}| \leq \Delta {\dot{I}} \,
\Delta{A}$, is a linear relationship between $\dot{I}$ and $A$
(SM.~\textsf{III}). A linear relationship between these variables implies the
probability distribution is of the form:
\begin{equation}
  p_x(t) = \frac{1}{Z(t)}\exp\left[-\int_{t_0}^{t}(c_1 a_x + c_2)dt'\right],
\end{equation}
with constants $c_1$, $c_2$, and partition function $Z(t)$. That is,
exponential probability distributions that are linear in the argument $a_x$
saturate the uncertainty relation, even when they are time dependent.

As one example of a distribution that saturates the bound, consider a system in
contact with a thermal bath of inverse temperature $\beta$. The equilibrium
distribution of a configuration $x$ is $p_x = e^{-\beta\epsilon_x}/Z$. If
temperature is varied over time such that the Boltzmann form of the
distribution is preserved, the uncertainty relations in the energy and entropy
representations,
\begin{align}
  \beta\dot{Q} &= -\beta\Delta \dot{I} \, \Delta \epsilon = \beta\dot\beta\Delta\epsilon^2\\
  \dot{S}/k_B &= -\Delta {\dot{I}} \, \Delta I= \beta\dot\beta\Delta\epsilon^2.
\end{align}
give $\dot{S}/k_B = \beta\dot{Q}$, the well known definition of the
thermodynamic entropy and the lower bound of the Clausius inequality for
reversible processes~\cite{Callen85}.

When the bound saturates, the evolution of the system is operating at the speed
limit. The intrinsic time for the observable is equal to the time scale set by
the Fisher information, i.e., $\tau = \speedtime$. For the specific case
considered here:
\begin{equation}
  \tau =\frac{1}{\Delta \dot I} = \frac{1}{|\dot{\beta}| \Delta \epsilon} = \frac{\Delta \epsilon}{|\dot Q|} =  \tau_{\scriptscriptstyle Q}.
  \label{eq:Quasi_tau}
\end{equation}
The time for the heat to evolve by one energy fluctuation is exactly the time
it takes the distribution to evolve to a distinguishable state. These time
scales are also equal to the intrinsic speed of the entropy
$\tau_{\scriptscriptstyle S} = k_B\Delta I/|\dot S| = \tau =
\tau_{\scriptscriptstyle Q}$. A quasistatic process is then one whose
thermodynamic time scales are equivalent to the statistical time scale $\tau$.
For this special driving protocol, the rate of change in the inverse
temperature is precisely the heat flow relative to the energy fluctuations:
$|\dot{\beta}| = |\dot Q|/\Delta \epsilon^2$. As we show next, though, even
complicated dynamics can nearly saturate the uncertainty bound.

\smallskip

\noindent{\textbf{\textsf{Model systems}}}\\
\noindent To illustrate our results, we numerically solved a model for
nonequilibrium self-assembly under periodic driving of the temperature, thermal
relaxation, and thermal annealing (SM.~\textsf{VII}). The analytical solution
of a two-state model is shown in SM.~\textsf{V}.

\begin{figure}
  \centering
  \includegraphics[width=.35\textwidth]{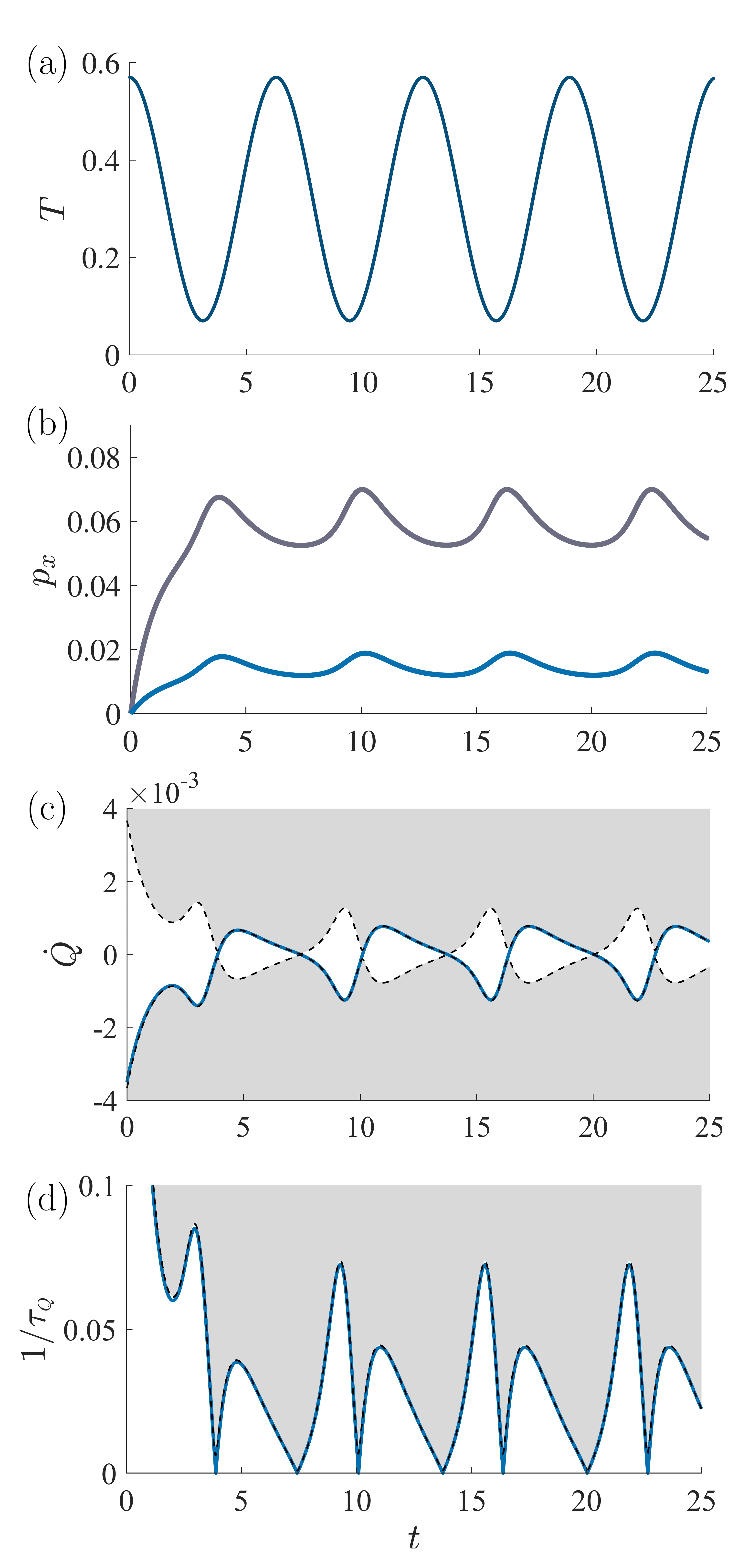}

  \caption{\label{fig:PerHeat} \textbf{Illustration of time-information
uncertainty relation and speed limit for a model of driven assembly of
monomeric units.} (a) Periodic protocol for the temperature. (b) Probability of
misassembled (grey) and optimally assembled (blue) states as a function of
time. (c) Heat flux as a function of time (blue) bounded by
$\pm\Delta\dot{I}\Delta\epsilon$ (dashed). (d) Speed of the heat exchanged as a
function of time (blue) is tightly bounded by the speed limit time set by the
Fisher information (dashed).}

\end{figure}

The self-assembly model we chose~\cite{grant2011analyzing} allows us to analyze
the energy exchanged as heat during an assembly process under arbitrary
protocols. The system can be found in three possible states $x$: dissociated
monomeric units, misbound aggregates, and an optimally bound configuration,
which will be denoted by $x_1$, $x_2$, and $x_3$, respectively. Initially the
system consists purely of monomers. As the system evolves, the temperature is
changed according to a given protocol and monomers transition into the
assembled states. Despite the simplicity of the model, the dynamics captures
the competition between kinetic trapping and binding strength, a phenomenon
also exhibited in more complicated models, such as those for chaperonin
proteins~\cite{grant2011analyzing}.

We take the dynamics to be the master equation, $\dot p(t) = \Omega(t)p(t)$.
The rate matrix $\Omega$ has non-negative off-diagonal elements and satisfies
$\Omega_{xx}(t) = -\sum_{x\neq y} \Omega_{xy}(t)$, which guarantees conservation
of probability. Its elements,
\begin{equation}
  \Omega = \begin{pmatrix}
  -c(M+1) & \alpha & \alpha^2 \\
  cM & -\alpha & 0\\
  c & 0 & -\alpha^2 
  \end{pmatrix},
\end{equation}
include a concentration-like variable $c = 0.02$, the number of possible
misbound states $M=5$, and $\alpha = \exp(-\epsilon_b/2T)$, a function of the
binding strength $\epsilon_b=0.1$ and the temperature $T$ with $k_B=1$. 

Consider a periodic variation of the temperature over time,
Fig.~(\ref{fig:PerHeat}a), with $T(t) = \gamma_1\cos(t) + \gamma_2$, with
$\gamma_1 = 0.25$ and $\gamma_2 = 0.32$ to keep the temperature in the range
used in Ref.~\cite{grant2011analyzing}. As a result of this driving protocol,
the probability of occupying the misbound state and optimally bound states also
oscillate in time, Fig.~(\ref{fig:PerHeat}b). The probability of observing the
monomer state also oscillates after a brief decay for the initial value of
$p_1(t_0) = 1$.

From our numerical solutions of this model, the product of the energy
fluctuations and intrinsic speed $\pm\Delta \dot{I}\Delta\epsilon =
\pm\Delta\epsilon/\tau$ (dashed line) provide upper and lower bounds to the
heat flux, $\dot{Q}$. The heat flux oscillates between, and closely follows,
the bounds, Fig.~(\ref{fig:PerHeat}c). The instantaneous speed of heat exchange
$\tau_{\scriptscriptstyle Q}$ during the driving protocol closely follows the
maximum $1/\tau$, except where the probabilities cross an inflection point,
Fig.~(\ref{fig:PerHeat}d). At these points where $\dot Q = 0$, the speed is
zero, while the limiting speed remains positive. The time-information
uncertainty relation and associated speed limit also hold when the system
undergoes thermal relaxation and thermal annealing, SM.~\textsf{VII}. The speed
limit for entropy is also confirmed by these calculations, SM.~\textsf{VI}.
Overall, this simple model illustrates that our uncertainty bounds apply to
thermodynamic observables under general nonequilibrium conditions.

\smallskip

\noindent{\textbf{\textsf{Conclusions}}}\\
\noindent According to thermodynamics, every natural process faces the
physical principle that structure formation or useful work production at a
particular speed comes at a cost: entropy production, energy dissipated as
heat, and wasted free energy. Here, we have shown that these thermodynamic
costs are restricted by fluctuations and satisfy a time-information
uncertainty relation. The mathematical form of this relation is similar to the
Mandelstam-Tamm version of the time-energy uncertainty relation, a significant
milestone in quantum mechanics. Because our formalism similarly requires few
details about the model system or the experimental conditions, we expect it to
be applicable to a broad range of physical and (bio)chemical systems. With no
assumption about the underlying model dynamics or external driving protocol, it
can also be applied to any nonequilibrium process with a differentiable
probability distribution. The uncertainty relations we derived for the flux of
heat, entropy (both its production and its flow), and the dissipated power
demonstrate that the time scale of their dynamical fluctuations away from
equilibrium are all bounded by the fluctuations in information rates. So, in
sum, while away from equilibrium, natural processes must also trade the
thermodynamic costs incurred for the speed of their evolution.

\acknowledgments

We acknowledge support from the U.S. Army Research Laboratory and the U.S.
Army Research Office under grant number W911NF-14-1-0359 and the National
Science Foundation under Grant No. 1856250. This work is further supported by
the John Templeton Foundation, UMass Boston Project No. P20150000029279, DOE
Grant No. DE-SC0019515, AFOSR MURI project ``Scalable Certification of Quantum
Computing Devices and Networks'', DoE ASCR Quantum Testbed Pathfinder program
(award No. DE-SC0019040), DoE BES QIS program (award No. DE-SC0019449), DoE
ASCR FAR-QC (award No. DE-SC0020312), NSF PFCQC program, AFOSR, ARO MURI, ARL
CDQI, and NSF PFC at JQI.

\vspace{-0.2in}

%


\section{Supplemental}

\noindent Sections \textbf{\textsf{\small I-IV}} contain proofs of results in the main
text. We suppress the time dependence in our notation (e.g., $p_x(t) = p_x$).
\vspace{2mm}

\noindent\textbf{\textsf{\small I.\@ Equation of motion:}} The surprisal rate
is defined as $\dot{I}_x = - d\ln p_x/dt$. Its mean is zero,
\begin{equation}
  -\langle\dot{I}\rangle = \sum_x^N p_x\frac{d\ln p_x}{dt} = \sum_x^N \frac{dp_x}{dt} = \frac{d}{dt}\sum_x^N p_x = 0,
\end{equation}
as a consequence of the conservation of probability, $\sum_x^Np_x = 1$. Using
these two facts, the equation of motion for the expectation value of an
observable is
\begin{eqnarray}\nonumber
  \frac{d}{dt}\langle A\rangle 
  &=& -\langle\dot{I}A\rangle + \left\langle\frac{dA}{dt}\right\rangle\\\nonumber
  &=& -\langle\dot{I}A\rangle + \langle\dot{I}\rangle\langle A\rangle + \left\langle\frac{dA}{dt}\right\rangle\\
  &=& -\operatorname{cov}(\dot{I},A) + \left\langle\frac{dA}{dt}\right\rangle
\end{eqnarray}
The final expression is Eq.~(1) in the main text. A covariance of zero
indicates two variables are uncorrelated. It does not necessarily mean that
they are statistically independent, since random variables that are
non-linearly related can also be uncorrelated.

\smallskip

\noindent\textbf{\textsf{\small II.\@ Entropy rate as covariance:}} The Shannon
entropy is the ensemble average
\begin{equation}
  S/k_B = -\sum_x^N p_{x} \ln p_{x} = \langle -\ln p\rangle=\langle I\rangle
\end{equation}
of the surprisal $I_x=-\ln p_x$. Using $\langle\dot{I}\rangle = 0$, its rate of
change,
\begin{align}\nonumber
  \frac{\dot{S}}{k_B} &= -\sum_{x}^N \dot{p}_{x} \ln p_{x} - \sum_{x}^N p_{x} \frac{d \ln p_{x}}{d t}\\\nonumber
  &= -\langle\dot{I}I\rangle + \langle\dot{I}\rangle\langle I\rangle\\
  &= -\operatorname{cov}(\dot{I},I),
\end{align}
can be expressed as the (negative) covariance of the surprisal and the
surprisal rate.

\smallskip

\noindent\textbf{\textsf{\small III.\@ Saturation of the uncertainty bound:}}
The covariance inequality for two random variables saturates when the random
variables are linearly related. For completeness, we show this for $\dot I$
and $A$. Consider the standardized variables,
\begin{align}
  \dot I_x' &= \frac{\dot I_x}{\Delta \dot I}, &  A'_x = \frac{A_x - \langle A \rangle}{\Delta A}. \nonumber
\end{align}
The expectation and standard deviation of both standardized variables is,
$\langle \dot I'\rangle =\langle A'\rangle= 0$ and $\Delta \dot I' = \Delta
A' = 1$. Defining the correlation as $\operatorname{cov}(X,Y)/\Delta X \Delta
Y$ and using the identity $\Delta (X - Y)^2 = \Delta(X)^2 + \Delta (Y)^2 -
2\operatorname{cov}(X,Y)$, we have
\begin{align}
  \Delta(\dot I' - A')^2 &= \Delta \dot I' + \Delta A' - 2\operatorname{cov}(\dot I',A')\nonumber \\
  &= 2[1-\rho(\dot I',A')]\nonumber \\
  &= 2[1-\rho(\dot I,A)].
\end{align}
The last line is a result of the fact that standardizing random variables does
not change the correlation $\rho(X',Y') = \rho(X,Y)$. Thus, the
condition $\rho(\dot I,A')=1$ is equivalent to $\Delta(\dot I' - A')^2 = 0$. A
zero variance means $\dot{I}_x' = A_x'$ with probability one. Taking the expectation of $\dot I'_x - A_x$, we see that $\langle \dot I'
- A'\rangle = 0$. As a result, $\dot I_x' = A_x'$ $\forall\, x$, or
\begin{align}
  \dot I_x = \frac{\Delta \dot I}{\Delta A} a_x - \frac{\Delta \dot I}{\Delta A} \langle A\rangle
  = c_1a_x -c_2.
\label{eq:LinRV}
\end{align} 
Using the definitions for $\tau_{\scriptscriptstyle{A}}$, (Eq. (5) in the main text) and $\tau := 1/\sqrt{I_F}$, saturating the bounds implies  that the time scales of the
system and observable are equal, i.e., $|\dot A| =
\Delta \dot I\Delta A \Rightarrow \tau = \tau_A$.

\noindent\textbf{\textsf{\small IV.\@ Bounds for a time-dependent Boltzmann
distribution:}} One example of a distribution that saturates the uncertainty
bound is a system in thermal contact with a reservoir that has a controllable
temperature $T(t)=T$ ($\beta = 1/k_BT$). The probability of being in state $x$
with energy $\epsilon_x$ is $p_x = e^{-\beta\epsilon_x}/Z$.

Here we show that in both the energy and entropy representation the bound is
saturated. The pertinent quantities are:
\begin{align}
\dot{p}_x &= -\left(\dot\beta\epsilon_x + \frac{\dot Z}{Z}\right)p_x, & I_x &= \beta\epsilon_x + \ln Z, \nonumber \\
\dot{I}_x &= \dot\beta\epsilon_x + \frac{\dot Z}{Z}, & S&= \beta U + \ln Z, \nonumber \\
\dot{Z} &= -\dot\beta\sum_x\epsilon_xe^{-\beta\epsilon_x}, & \frac{\dot Z}{Z} &=-\dot\beta U.
\end{align}
Let us start with the heat flux,
\begin{align}
\dot Q &= \sum_x \dot p_x\delta \epsilon_x \nonumber \\
&= -\sum_x\left(\dot\beta p_x \epsilon_x + p_x\frac{\dot Z}{Z}\right) \nonumber \\
&=-\dot\beta\left(\sum_x p_x\epsilon_x^2 - U^2\right)\nonumber \\
&=-\dot\beta\Delta\epsilon^2.
\label{eq:Qdot}
\end{align}
The covariance of $\dot{I}$ and $\epsilon$ reads
\begin{align}
\operatorname{cov}(\dot I,\epsilon) &= \sum_xp_x\dot{I}\delta\epsilon_x\nonumber \\
&= \sum_x p_x\dot{I}\epsilon_x - \sum_x p_x\dot{I}_xU\nonumber \\
&= \sum_x p_x\epsilon_x\left(\dot\beta\epsilon_x +\frac{\dot Z}{Z}\right)\nonumber \\
&= \dot{\beta}\left( \sum_x p_x \epsilon_x^2 - U\sum_x p_x\epsilon_x\right)\\
&= \dot\beta\Delta\epsilon^2,
\end{align}
which is equal and of opposite sign to $\dot{Q}$ as claimed in the main text.
We can arrive at the same result using Eq.~(\ref{eq:LinRV}). Starting with $c_1
= \Delta \dot I/ \Delta A = \sqrt{I_F}/\Delta A$, the Fisher information equals
\begin{align}
I_F &= \sum_x \frac{\dot p_x^2}{p_x} \nonumber \\
&= \sum_x \frac{\left(\dot \beta \epsilon_x +\frac{\dot Z}{Z}\right)^2p_x^2}{p_x}\nonumber \\
&= \dot\beta^2\sum_x p_x\left(\epsilon_x - U\right)^2\nonumber \\
&= \dot \beta^2\Delta \epsilon^2.
\end{align}
The second coefficient is $c_2 = \sqrt{I_F}U/\Delta \epsilon$, making the
change in information $\dot I_x = \dot\beta\epsilon_x - \dot Z/Z$. The
covariance with $\epsilon$ follows as above to again give
$\operatorname{cov}(\dot I,\epsilon) = \dot \beta\Delta\epsilon^2$.

Next, looking at $\dot{S}$, we see that for this exponential distribution
\begin{align}
\dot{S}/k_B &= \sum_x \dot{p}_x\delta I_x\nonumber \\
&= \sum_x \dot p_x\left(\beta\epsilon_x + \ln Z\right) \nonumber \\
&= -\beta\sum_x \dot p_x\epsilon_x \nonumber \\
&= \beta\dot Q,
\end{align}
recovers the equilibrium relationship between entropy and heat. Using
Eq.~(\ref{eq:Qdot}) we also find that the change in entropy can be written as,
$\dot S = -\beta\dot\beta\sigma_\epsilon^2$. The covariance between $\dot{I}$
and $I$,
\begin{align}
\operatorname{cov}(\dot{I},I) &= \sum_x p_x\dot{I}(I_x - S)\nonumber \\
&= \beta\sum_x p_x\dot{I}_x\epsilon_x + \sum_x p_x\dot{I}_x\ln Z - \sum_x p_x\dot{I}_x S\nonumber \\
&= \beta\sum_x p_x\dot{I}_x\epsilon_x\nonumber \\
&= \beta\sum_x p_x\epsilon_x\left(\dot\beta\epsilon_x +\frac{\dot Z}{Z}\right)\nonumber \\
&= \beta\dot\beta\Delta\epsilon^2,
\end{align}
confirms that $\dot{S} = -\operatorname{cov}(\dot{I},I)$. Finally, looking at
the right hand side of the entropic uncertainty relation, we have the
fluctuations in the surprisal
\begin{align}
  \Delta I^2 &= \sum_x p_x(I -S)^2 \nonumber \\
  &= \sum_x p_x(\beta\epsilon_x +\ln Z - \beta U - \ln Z)^2 \nonumber \\
  &=\beta^2\Delta\epsilon^2,
\end{align}
and the surprisal rate
\begin{align}
\Delta{\dot{I}}^2 &= \sum_x p_x\dot{I}^2 \nonumber \\\nonumber
&= \sum_x p_x\left(\dot\beta\epsilon_x + \frac{\dot Z}{Z}\right)^2 \\\nonumber
&=\dot\beta^2(\langle \epsilon^2\rangle - U^2) \nonumber \\
&= \dot\beta^2\Delta\epsilon^2.
\end{align}
Putting these results together, we find that 
\begin{align}
  -\dot Q &= \dot\beta\Delta \epsilon^2 = \operatorname{cov}(\dot{I},\epsilon) = \Delta {\dot{I}}\Delta \epsilon, \nonumber \\
  -\dot S &= \beta\dot\beta\Delta \epsilon^2 = \operatorname{cov}(\dot{I},I) = \Delta {\dot{I}}\Delta I.
\end{align}

\noindent Sections \textbf{\textsf{\small V-VII}} contain additional results for model
systems. 

\begin{figure}[t]
\includegraphics[width=\columnwidth]{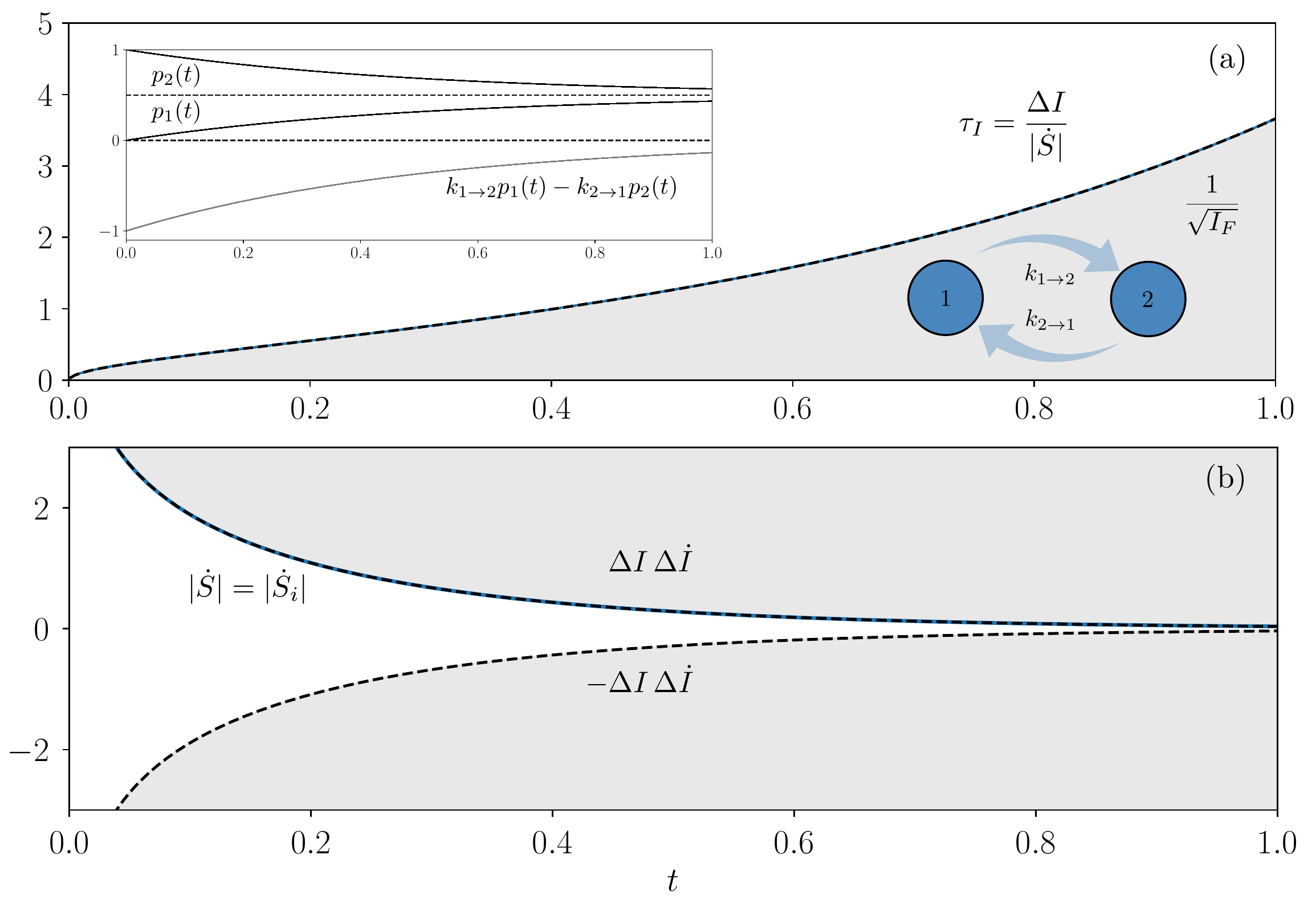}

\caption{\label{fig:two-state}Irreversible relaxation of a two-state system to
equilibrium. The intrinsic dynamical time scale $1/\sqrt{I_F}$ sets a limit
(dashed) on the characteristic time scale of the entropy rate/production
$\dot{S}=\dot{S}_i$ (solid blue). We use $k_{12}=k_{21}=1$ so there is no
entropy flow, $\dot{S}_e =0$. As the system nears equilibrium, the time to
reach new distinguishable states diverges and the speed goes to zero
$1/\tau_I\to 0$. Inset: Relaxation of probability of each state (black) from
$p_2=1$, $p_1(0)=0$ and the current (gray) showing that the system is not at
equilibrium.}

\end{figure}

\noindent\textbf{\textsf{\small V.\@ Two-state system:}} Let us analytically
solve a simple two state system governed by master equation dynamics,
\begin{eqnarray}
  \frac{d p_1}{dt} &=& k_{21} p_2 - k_{12} p_1\\
  \frac{d p_2}{dt} &=& k_{12} p_1 - k_{21} p_1,
\end{eqnarray}
with rate coefficients $k_{12} = k_{1\to 2}$ and $k_{21} = k_{2\to 1}$. For
time-independent $k_{ij}$, the solutions are:
\begin{eqnarray}
  p_1 &=& \frac{k_{21}}{k}\left[1-e^{-k t}\right] + e^{-k t} p_1(0)\\
  p_2 &=& 1-p_1.
\end{eqnarray}
where $k = k_{21}+k_{12}$. In the limit $t\to\infty$, the probabilities of each
state reach steady values $p_1^\text{eq} = k_{21}/k$ and $p_2^\text{eq} =
k_{12}/k$. Together with the time derivatives,
\begin{equation}
  \frac{dp_1}{dt} = -\frac{dp_2}{dt} = e^{-kt}\left[k_{21}-k p_1(0)\right]
\end{equation}
we constructed the Fisher information, entropy rate, and fluctuations in
surprisal rate. These quantities are shown as functions of time in
Fig.~\ref{fig:two-state} with $k_{12}=k_{21}=1$ and initial conditions
$p_1(0)=0$, $p_2(0)=1$. The Fisher information is
\begin{eqnarray}
  \Delta \dot{I} = I_F &=& \frac{(\dot{p}_1)^2}{p_1}+\frac{(\dot{p}_2)^2}{p_2}
  = \frac{(\dot{p}_1)^2}{p_1(1-p_1)}.
\end{eqnarray}
The entropy rate reads
\begin{equation}
  \dot{S}/k_B = -\dot{p}_1\ln p_1 - \dot{p}_2\ln p_2.
\end{equation}
The surprisal variance is
\begin{equation}
  \Delta I = p_1(\ln p_1)^2 + p_2(\ln p_2)^2 - S^2.
\end{equation}
As shown in the figure,
\begin{equation}
  \frac{\Delta I}{|\dot{S}|/k_B} \leq \frac{1}{\sqrt{I_F}}.
\end{equation}
Since $k_{12}=k_{21}$ in this example, the entropy flow $\dot{S}_e = 0$ and the
speed limit is imposed on the entropy production: $\dot{S}=\dot{S}_i$.
\begin{figure}
\centering
\includegraphics[width=.45\textwidth]{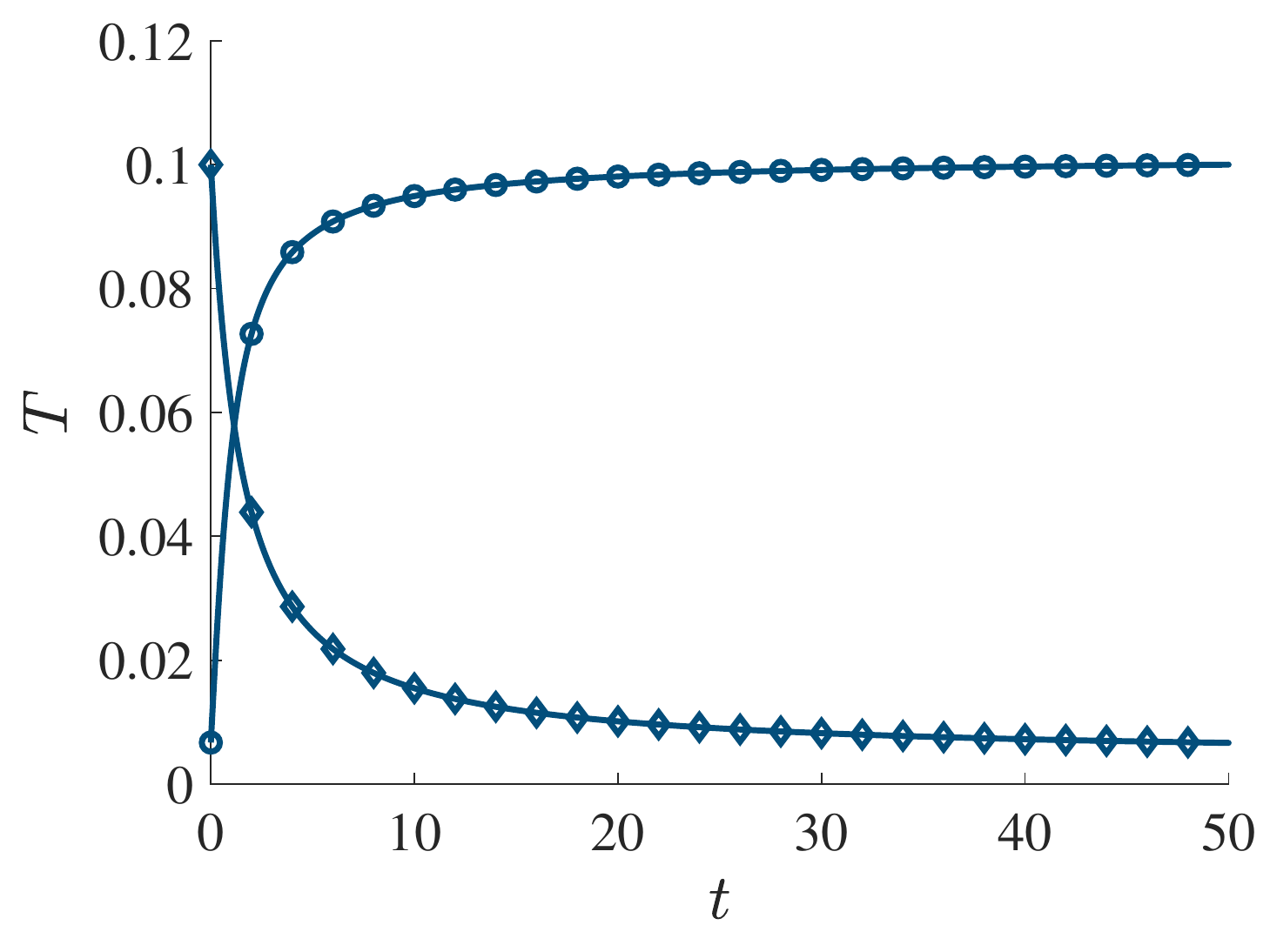}

\caption{\label{fig:TempProt} Temperature profiles used for annealing and
heating. The annealing protocol (diamonds) goes from $T(0) = 0.1$ to $T(50) =
0.006$, while the heating protocol (circles) goes from $T(0) = 0.006$ to
$0.1$.}

\end{figure}

\smallskip

\noindent\textbf{\textsf{\small VI.\@  Entropy representation for periodically driven assembly:}}
\begin{figure}[!t]
\centering
\includegraphics[width=.9\columnwidth]{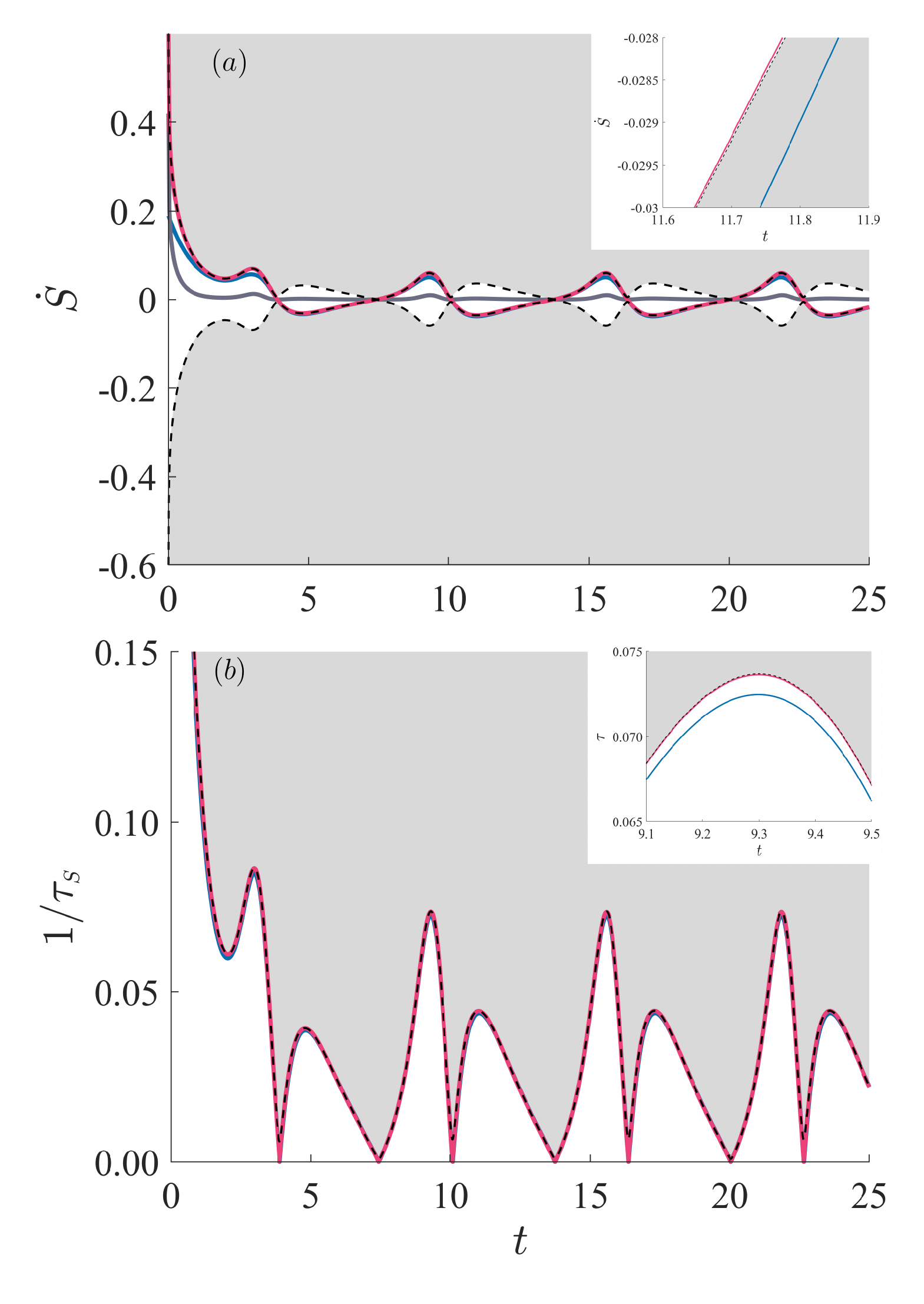}

\caption{\label{fig:Ent_per} (a) The change in entropy red line with its
constituent parts, the entropy flow (blue line) and production (dark grey
line). $\dot S$ closely follows the upper and lower bounds set by the
uncertainty relation. The inset shows that while $\dot S$ always lies within
the bounds the entropy flow can deviate. (b) the speed is close to the upper
bound set by the Fisher information. The inset shows the speed for
$S$ in red and $Q$ in
blue. For this example, $1/\tau_{\scriptstyle S} \geq
1/\tau_{\scriptstyle Q}$.}

\end{figure}
In the entropy representation, our time-information uncertainty relation is:
\begin{equation}
|\dot S| \leq \Delta \dot I \Delta I.
\end{equation}
We confirm this result using the same assembly model and periodic protocol for
the temperature as in the main text, Eq.~(\ref{fig:TempProt}). The upper
and lower bounds on $\dot S$ are shown in Fig.~(\ref{fig:Ent_per}a). The
entropy production (grey) and flow (blue) are also shown. The entropy rate
switches between the positive and negative bounds over the time. The entropic
speed $1/\tau_{\scriptscriptstyle S}$ also closely follows the maximum speed
allowed by the fluctuations in the surprisal rate. Since both
$\tau_{\scriptscriptstyle S}^{-1}$ and $\tau_{\scriptscriptstyle Q}^{-1}$ are
bound by $1/\tau$, we can compare the speed of both representations. The inset
to Fig. (\ref{fig:Ent_per}b) shows that $\tau_{\scriptscriptstyle Q}^{-1}$
(blue line) lies farther from the maximum speed than $\tau_{\scriptscriptstyle
S}^{-1}$. 

\newpage

\noindent\textbf{\textsf{\small VII.\@  Annealing and heating of assembly process:}}
\begin{figure*}
\centering
\includegraphics[width=.9\textwidth]{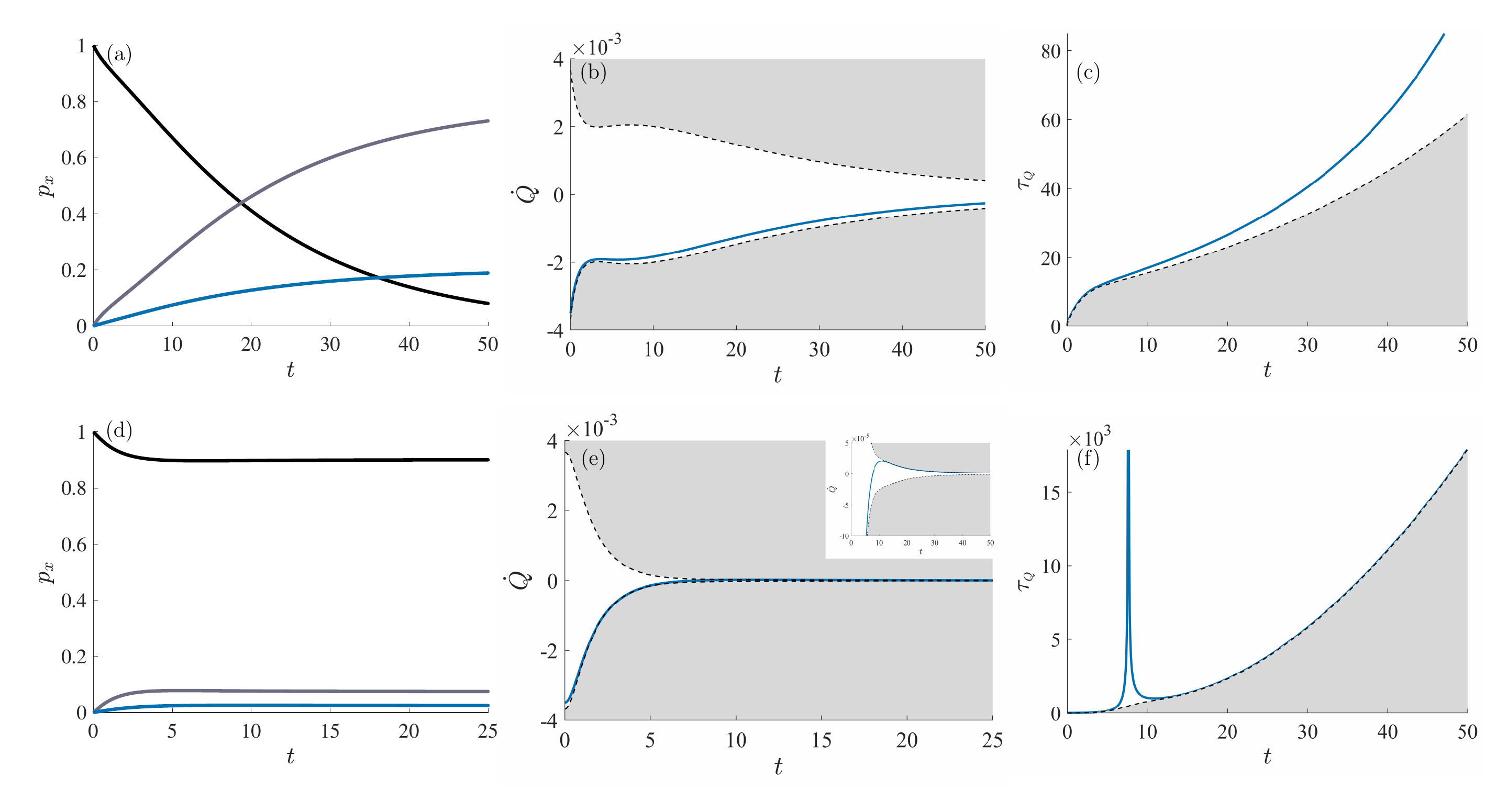}

\caption{Probability of the monomer (black), misbound states (grey), and
optimally bound (blue) as a function of time for the (a) annealing protocol and
(d) heating protocol. (b) The uncertainty relation bounds (dashed lines) on the
heat flux, $\dot Q$.  (c) Both the speed limit time (dashed line) and the
intrinsic time $\tau_{\scriptscriptstyle Q}$ grow as the system approaches
equilibrium. (e) Using the heating protocol, the heat flux and uncertainty
relation bounds are path dependent. The inset shows that $\dot Q$ crosses zero.
(f) At $\dot Q = 0$, the time scale $\tau_{\scriptscriptstyle Q}$
diverges.}

\label{fig:Qdot_cool}
\end{figure*}

In addition to the periodic driving protocol used in the main text, we consider $\dot Q$ for both a heating and an annealing protocol. Given an initial
temperature $T(t_0)$, we cool the system according to $T_c(t) = 2T(0) +
\gamma_1\atan(\gamma_2 - t)$, where we use $\gamma_1 = 0.1246$ and $\gamma_2 =
-1.033$, which gives $T(0) = 0.1$ and $T(50) = 0.006$. To heat the assembling
system, we swap the initial and final temperatures used in the annealing
protocol: $T_h(t) = \gamma_1\atan(\gamma_2 +t)$, where now $\gamma_1  = 0.0645$
and $\gamma_2 = 0.1033$. Both temperature protocols are shown in
Fig.~(\ref{fig:TempProt}).

The probability of being in the monomer state (black line), misbound state
(grey line) and optimally bound state (blue line) is shown in
Fig.~(\ref{fig:Qdot_cool}a). The heat flux $\dot Q$ is negative throughout and
moves farther from the minimum bound. Panel (c) shows the time
$\tau_{\scriptscriptstyle Q}$ instead of the speed $1/\tau_{\scriptscriptstyle Q}$, which begins
near the speed limit time but deviates as the system approaches equilibrium.

Heating the system according to the second temperature protocol does not
produce the same change in heat or the same bounds. This path dependent nature
of heat flow is illustrated in Fig.~(\ref{fig:Qdot_cool}e). Over time, the heat
flux switches from negative to positive before being heavily constrained by the uncertainty bounds approaching zero as the distribution approaches a steady state. If we look at the minimum characteristic time $\tau_{\scriptscriptstyle Q}$ in
panel (f), we see it diverges as $\dot Q$ crosses zero but otherwise closely
follows the lower bound $1/I_F$.

\end{document}